\documentclass[12pt]{iopart}

\begin{document}

\title[The high energy frontier of A-A]{The high energy frontier of ultrarelativistic nucleus-nucleus collisions}

\author{Federico Antinori}

\address{INFN, Sezione di Padova, via Marzolo 8, I-35131 Padova, Italy}
\ead{federico.antinori@pd.infn.it}
\begin{abstract}
An outlook on physics at the high energy frontier of nucleus-nucleus collisions is presented, on the basis of the new results presented at Quark Matter 2011 by the LHC and RHIC experiments.
\end{abstract}


\section{Introduction}

I have been asked by the organizers of Quark Matter 2011 to provide a perspective on the high energy frontier of nucleus-nucleus collisions in the wake of this historic edition of the conference: the first after the start of operation of the Large Hadron Collider (LHC).
The usual subjectivity disclaimers customary for this sort of reviews apply here, more than ever, given the plenty of results to chose from in this first LHC harvest.
"High energy" has been defined for me as the top energy of the Relativistic Heavy Ion Collider (RHIC) and above. The RHIC energy scan programme, in particular, is out of the scope of this paper and is covered in the contribution of Marcus Bleicher \cite{bleicher}.
I have -- somewhat arbitrarily -- divided the presently known landscape of high energy nuclear collisions into the following regions: global observables (section 2), correlations (section 3), quarkonia (section 4), high $p_T$ suppression (section 5), identified particles (section 6), jets (section 7) and heavy flavours (section 8). For each of these areas I have tried to glimpse towards the {\it terrae incognitae} that lay at the frontier of our current knowledge.
I have drawn my conclusions in section 9.

\section{Global observables}

The multiplicity of produced particles is probably the most basic of observables. It provides information on the energy density achieved in the collisions and constitutes a primary input for most model calculations.
For central Pb-Pb collisions at the LHC (at $\sqrt{s_{NN}}=2.76$ TeV),
about 8 charged particles per unit of pseudorapidity $\eta$ are produced\cite{mult1,alberica}: about twice as many as at RHIC.
This is larger than most of the predictions and about 50 \% more than expected from simple phenomenological extrapolations from RHIC energy: the logarithmic law that described the energy dependence of multiplicities up to RHIC has finally broken down. The average amount of transverse energy produced per unit of pseudorapidity per participant pair in central collisions is about 9 GeV, or about a factor 2.7 larger than at RHIC (the larger multiplicity at LHC being accompanied by an increase in the average transverse momentum of the produced particles), corresponding to an energy density of about 16 GeV/fm$^3$ (taking the conventional value of 1 fm/$c$ for the plasma formation time).
The centrality dependence of the charged particle multiplicity  is rather mild, favouring models incorporating some mechanism (such as parton saturation) moderating the increase with centrality of the average multiplicity per participant pair\cite{mult2,constantin,steinberg}.
Once the multiplicities are rescaled to account for the difference in the central values, the centrality dependences at the LHC and at RHIC are remarkably similar.

Hanbury Brown -- Twiss (HBT) interferometry \cite{HBT1,HBT2} exploits the quantum intereference between identical bosons (e.g. charged pions) to evaluate the size of the system at the time of decoupling. Compared to RHIC, the ALICE experiment finds an increase in the dimensions of the system in all three components \cite{BEpaper,adam} (including, finally, also the sidewards component $R_{side}$). The system expands significantly more than at RHIC, with an estimated increase by about a factor 2 of the homogeneity volume for central collisions.

In the area of global observables, the frontier is presently at the study of Colour Glass Condensate/saturation effects, with the measurement of the pseudorapidity dependence of the multiplicity, and -- hopefully soon -- the study of proton-nucleus collisions. On the HBT front, the next step is to perform identified-particle interferometry, to establish how the size of the emitting source depends on the particle flavour and mass.

\section{Correlations}

Particle correlations took centre-stage at this conference. The average value of the {\it elliptic flow} coefficient $v_2$ (the second coefficient in the Fourier expansion of the azimuthal distribution) continues to be large at the LHC (roughly 20\% larger than at RHIC for semi-central events)\cite{v2,raimond,jia,velkovska}: the behaviour of the system is still very close to that of an ideal liquid. When measured as a function of transverse momentum, the values of $v_2$ at the two colliders are close: although the events at the LHC are on average harder than those at RHIC, the hydrodynamical properties of the system at the two energies seem to be rather similar. The $v_2$ coefficient is still significantly different from zero at high transverse momenta (measurements are available out to 20 GeV/$c$ \cite{dobrin,trzupek}), providing important constraints for the energy loss models. Fluctuations have been one of the highlights of this conference. For symmetric nuclei, the azimuthal distribution of the participant nucleons should contain only even terms, and is expected to be dominated by the second (elliptic) Fourier coefficient, reflecting the elongated distribution of participants in the transverse plane for non-central collisions. In the presence of event-by-event fluctuations in the positions of the individual nucleons, however, odd terms can appear in the participants' distribution, leading, via hydrodynamic expansion, to higher order harmonics in the momentum distributions of the emitted particles (see \cite{alverroland} and, for a discussion at this conference, \cite{luzum}). The third order harmonics ($v_3$), quantifying the {\it triangular flow}, is the next highest term in the Fourier decomposition of the azimuthal distribution. An odd term, it is exclusively due to initial state fluctuations. When calculated with respect to the event plane determined from the $v_2$ analysis, $v_3$ is negligible, as would be expected for participants fluctuations \cite{raimond,hiarm}. While it is significantly smaller than $v_2$ for non-central collisions, $v_3$ has a much weaker centrality dependence. For ultra-central collisions (0-1\%) at the LHC, the two terms actually become comparable in magnitude, giving rise to a prominent long-$\eta$-range triangular structure in the two-particle azimuthal correlation function, visible without the need for any $v_2$ subtraction, with a main peak around $\Delta \varphi = 0$, corresponding to the long-range correlation labelled the {\it ridge} at RHIC, and two secondary peaks around $\Delta \varphi = 120^{\circ}$ and $\Delta \varphi = 240^{\circ}$, corresponding to the position of the RHIC {\it Mach cone} \cite{raimond,jia,hiarm,andrew, li}. For correlations involving particles with transverse momenta in the range of a few GeV/$c$, the first five harmonics seem to be sufficient to describe the observed long-range azimuthal correlations, and both the ridge and the Mach cone, for which explanations invoking medium response effects were proposed, seem to find a natural explanation in terms of initial state fluctuations. Results consistent with this picture have also been shown by the RHIC experiments \cite{esumi,sorensen}.
Inasmuch as azimuthal correlations are driven by the global event anisotropy, as expected for a collective hydrodynamic expansion, the Fourier components of the decomposition of the two-particle correlation should be given by the product of the single particle coefficients. This is indeed found to be the case, out to transverse momenta of the order of 3-4 GeV/$c$ for $v_2$ up to $v_5$ \cite{jia, andrew, janfiete}. Such a factorization, instead, does not hold for $v_1$ (which is expected to be substantially influenced by momentum conservation effects).

This is an alluring frontier: the spectral decomposition of the azimuthal correlation (already dubbed "the Annecy spectrum") promises to provide a beautiful tool for comparison with theory on the way to the extraction of quantitative information on the initial conditions and the medium viscosity. By the way, it would be very interesting to study the Fourier decomposition of the correlations around the ridge structure observed by CMS in high multiplicity proton-proton collisions \cite{ppridge}. It will also be essential to understand if there is any additional room for medium response effects, by looking at the "small print" on the away side, two-dimensionally in $\eta - \varphi$, and using, whenever possible, the information on the pseudorapidity of the recoiling parton (for instance around re-emerging away-side jets or around reconstructed recoiling heavy flavour particles).

\section{Quarkonia}

Several new results on quarkonia were presented at this conference. The PHENIX collaboration has presented an update on J/$\psi$ suppression using the 2007 data sample \cite{newphenixjpsi,dasilva}. The previous results, and in particular the stronger suppression at forward than at central rapidity, are confirmed, with reduced statistical and systematic uncertainties. STAR have shown results on the J/$\psi$ $v_2$ \cite{masui}, which is found to be compatible with zero up to a transverse momentum $p_T \simeq$ 8 GeV/$c$ for semi-central collisions, somewhat disfavouring the coalescence of thermalised charm quarks. The first results on J/$\psi$ suppression at the LHC have also been presented. For $p_T > 0$ and $2.5 < y < 4$, ALICE observes a J/$\psi$ nuclear modification factor $R_{AA}^{J/\psi}$ of about 0.5, practically independent of centrality\cite{gines}, while for $p_T > 6.5$ GeV/$c$ and central rapidity CMS reports substantial centrality dependence and larger suppression, with $R_{AA}^{J/\psi}$ as low as 0.2 for central collisions\cite{silvestre}. ATLAS also observes strong centrality dependence in the central-to-peripheral nuclear modification factor $R_{cp}^{J/\psi}$ at high $p_T$\cite{atlasjpsi}.

STAR presented preliminary results on the production of $\Upsilon$(1S+2S+3S) at RHIC \cite{masui}, reporting substantial suppression, about a factor 3, in the nuclear modification factor for the combined yield. CMS reported suppression for the $\Upsilon$(1S) (around 0.6, with a weak energy dependence) and further suppression by about a factor 3 relative to the $\Upsilon$(1S) for the excited states $\Upsilon$(2S+3S)\cite{silvestre}.

At the quarkonium frontier, we should not be far from being able to see the detailed pattern of quarkonia suppression (and perhaps recombination) in full glory,
with high statistics measurements and full control of the heavy flavour baselines (and, for the J/$\psi$, contaminations). Establishing a good pA reference will also be essential in order to disentangle the contributions from cold nuclear effects. Such measurements should allow us to put strong constrains on the value of the initial temperature and on the hadronisation mechanisms.

\section{High $p_T$ suppression}

Results on high $p_T$ suppression at the LHC have been presented by the three experiments ($R_{AA}$ by ALICE \cite{RAA,harry} and CMS \cite{lee} and $R_{cp}$ by ATLAS \cite{milov}). The production of high $p_T$ particles in Pb--Pb collisions at the LHC is strongly suppressed. $R_{AA}$ reaches a minimum of about 0.14 for transverse momenta around 6-7 GeV/$c$, and then increases again very slowly, almost levelling off around 0.5 for transverse momenta above 30 GeV/$c$.
The next step here is the study of the suppression as a function of the orientation with respect to the event plane
(preliminary results on this have been presented by ALICE\cite{dobrin}). Such measurements should provide information on the path length dependence  ($L^2, L^3, ...$) of the energy loss. An example of such an analysis has been shown by PHENIX, who found that the centrality dependences of the azimuthal asymmetry and of the nuclear modification factor can be reproduced simultaneously assuming a cubic path length dependence for the energy loss, as suggested by AdS/CFT calculations\cite{bathe}.
Further measurements both at RHIC, where asymmetric collisions (Cu--Au, U--U) are being planned, and at the LHC, should allow putting tight additional constraints on the energy loss models.

\section{Identified particles}

Results on identified particle spectra at the LHC have been presented by the ALICE collaboration \cite{michele}. While pion and kaon $p_T$ spectra are in good agreement with hydrodynamic model predictions \cite{uli}, the proton spectra are not reproduced by the model, neither in the yield (overpredicted by the model calculations) nor in the shape (harder in the data than in the model). The measured p/$\pi$ ratio (around 0.05), is at odds with thermal models predictions obtained assuming temperatures in the 160-170 MeV range (\cite{cleymans,pbmetal}).
Hydrodynamics also has trouble reproducing the $v_2$ of protons for central collisions, which is lower than expected from hydro calculations in the 1-2 GeV range \cite{raimond}, while the predictions for the $p_T$-differential $v_2$ for kaons and pions are in good agreement with the data.

Understanding the hydrodynamics of protons is clearly one of the highest priorities now. It is also essential to investigate the extent of disagreement between the data and the thermal predictions for the particle rates, with comprehensive, high statistics fits including strange and multi-strange particles out to the $\phi$, $\Xi^-$ and $\Omega^-$.

\section{Jets}

Intriguing results on jet production at the LHC have been shown by ATLAS\cite{cole} (using the anti-$k_T$ algorithm \cite{cacciari}) and CMS\cite{roland} (based on the use of an iterative cone algorithm \cite{kodolova}). A strong increase in the di-jet unbalance is observed when comparing Pb-Pb to pp collisions\cite{atlasjet, cmsjet}, when jets are studied with jet radii $R=\sqrt{(\Delta\eta)^2+(\Delta\varphi)^2}\simeq 0.2 - 0.4$. On the other hand, the fragmentation functions are found to be very similar to the proton--proton ones, and there is no visible decorrelation in $\varphi$ between the two jets recoiling against each other: the distribution of the azimuthal distance between the leading and the subleading jet in Pb--Pb collisions is very similar to the pp one. The energy balance seems to be restored by a large number of low transverse momentum (mostly $p_T < 2$ GeV/$c$) particles emitted at larger cone values\cite{cmsjet}. The suppression of jets is substantial and apparently rather independent of the jet energy, with a nuclear modification factor ($R_{cp}$) around 0.5 for central (10\%) events, flat out to 250 GeV or so of transverse energy.

From RHIC, the STAR experiment reports evidence for broadening and softening of jets in central Au--Au collisions \cite{helen}. Using both the $k_T$\cite{kt} and anti-$k_T$ clustering algorithms, they observe a substantial decrease in the jet yield when the radius is reduced from 0.4 to 0.2 (see also \cite{mateusz09}). When investigating the energy balance with jet-hadron correlations, they observe a deficit of energy with respect to pp collisions at associated hadron transverse momenta $p_T^{assoc}>2$ GeV/$c$, approximately compensated by an excess at $p_T^{assoc}<2$ GeV/$c$\cite{ohlson}. PHENIX, reconstructing jets using a gaussian filter, report a centrality-dependent suppression of the jet yields (up to about a factor 2 relative to pp for central Au--Au collisions)\cite{purschke}. Interestingly, they also report non-negligible jet suppression in d--Au collisions (about a 20\% effect in central versus peripheral d--Au), underlying once more the importance of measuring the "cold nuclear reference".

This is certainly one of the most lively (and puzzling) areas in our field at the moment. The prospect are enticing, both in the high-energy sector, with $\gamma$-jet and $Z^0$-jet fragmentation function measurements coming up, and in the softer one, with the exploration of the jet surroundings at low transverse momenta (in search for signs of broadening, softening, reheating, ...) and possibly with the extension to more "organic" jet clustering algorithms (could one somehow try to catch the whole modified jet, and then look inside as a function of the radial distance?) It would also be very useful to reach an agreement among the experiments on the set of standard clustering algorithms to be used systematically by all, in order to enable fully quantitative comparisons.
Jet hadro-chemistry studies, with the measurement of identified particle fragmentation functions, should also improve our understanding of in-medium fragmentation, and the use of b-tagged jets should allow investigating the differences in the fragmentation between quarks and gluons.
It would also be very interesting to single out and study in fine detail the most extreme "mono-jets" events, where one of the jets is almost completely absorbed.

\section{Heavy flavours}

Another busy frontier area in the field of ultra-relativistic nucleus-nucleus collisions is the study of the production of heavy flavours. There has been partial progress since Quark Matter 2009 on the PHENIX/STAR discrepancy on heavy flavour production, with new results from STAR on the non-photonic electrons' $p_T$ spectrum in pp collisions\cite{xie}, now in agreement with those of PHENIX. The STAR experiment has presented a D$^0$ signal from Au--Au collisions obtained by combining K$^-\pi^+$ track pairs from the TPC.
The have measured the nuclear modification factor $R_{AA}$ for the D$^0$, out to $p_T ~ 3-4$ GeV/$c$, and found it to be compatible with 1 for the 0-80\% centrality range \cite{zhang}. Both STAR and PHENIX are planning to upgrade the apparatus with the addition of vertex detectors \cite{gagliardi, bathe}, which should allow them to separate the vertices of the weak decays of heavy flavour particles from the primary vertex and lead to significant improvements in their heavy flavour capability. Heavy flavour vertexing is already available in the LHC experiments. Reconstructed D$^0$ and D$^+$ signals have been presented by ALICE \cite{andread, andrear}, who observe substantial suppression of the production of D mesons over the currently available transverse momentum range (out to 12 GeV/$c$) in both central (0-20\%) and peripheral (40-80\%) Pb--Pb collisions. For both centrality classes the values of the nuclear modification factor $R_{AA}$ are found to be compatible with those of the pions, with perhaps a hint of a lower suppression for D than for pions only below 5 GeV/$c$ or so. This is somewhat puzzling, since charm hadrons should originate from the hadronisation of (charm) quarks, while pions at such transverse momenta originate mostly from gluons, which are expected to have a weaker coupling to the medium than quarks, and therefore to lose less energy (for instance, in BDMPS energy loss calculations for LHC, D are expected to be suppressed about a factor 2 less than pions for $p_T \sim 8$ GeV/$c$ \cite{adsw}).
Could charm quarks be essentially thermalised in the QGP, and thereby "lose memory" of their energy loss history?
ALICE have also presented results on the nuclear modification factor of heavy flavour electrons at rapidity $y \sim 0$ and $1.5 < p_T < 6$ GeV/$c$ \cite{silvia} and muons at $y \sim 3$ and $4 < p_T < 10$ GeV/$c$ \cite{xiaoming}, with good agreement in the overlapping region of transverse momenta. The production of muons is found to be suppressed by about a factor 3 in the 5-10 GeV/$c$ $p_T$ range.
CMS have presented results on the nuclear modification factor of J/$\psi$ originating in the decay of B mesons for $p_T^{(J/\psi)} > 6.5$ GeV/$c$\cite{silvestre}. The suppression is again about a factor 3, with very little -- if any -- centrality dependence.

The future will bring high statistics D and B measurements, allowing us to establish the mass and colour charge dependence of the parton energy loss. The measurement of the azimuthal asymmetry coefficient $v_2$ for D mesons will provide an important test of the degree of thermalisation of charm. The subtraction of the D decay background from the heavy flavour electron spectra will allow the extraction of a pure beauty electron spectrum, enabling the measurement of beauty energy loss in a wide transverse momentum range. Beauty-tagging of jets will allow measuring the in-medium fragmentation of b quarks.

\section{Conclusions}

With the start of LHC operations and the upcoming upgrades of the RHIC experiments, high energy nucleus-nucleus collisions are entering an era of precision measurements that should allow us to impose tight constraints on the properties of the medium.

Many ideas that were taken more or less for granted at the time of Quark Matter 2009 are now being seriously questioned: the expression "death of the Mach cone and ridge" was often heard in the corridors, and we are now having second thoughts on issues such as thermal particles yields and identified-particle hydrodynamics. Our understanding of parton energy loss is challenged by the LHC results on jet and heavy flavour energy loss.

The outlook is brighter than ever: with a high luminosity LHC Pb--Pb run (5 - 10 times the 2010 luminosity) coming up later this year and possibly a p--Pb run as early as next year, we could be in for some paradigm shifting at Quark Matter 2012!

\section*{Acknowledgments}

I would like to thank Peter Braun-Munzinger, Brian Cole, Andrea Dainese, David d'Enterria, Paolo Giubellino, Barbara Jacak, Peter Jacobs, Karel \v{S}afa\v{r}\'ik, Carlos Salgado, J\"urgen Schukraft, Edward Shuryak, Peter Steinberg, Xin-Nian Wang, Urs Wiedemann and Nu Xu for their input during the preparation of this work, and the organisers for the honour and pleasure of presenting the final talk at this memorable edition of Quark Matter.

\end{document}